\documentclass[%
 aip,
 amsmath,amssymb,
 reprint,%
]{revtex4-1}

\usepackage{graphicx}
\usepackage{dcolumn}
\usepackage{bm}

\usepackage[utf8]{inputenc}
\usepackage[T1]{fontenc}
\usepackage{mathptmx}
\usepackage{etoolbox}

\makeatletter
\def\@email#1#2{%
 \endgroup
 \patchcmd{\titleblock@produce}
  {\frontmatter@RRAPformat}
  {\frontmatter@RRAPformat{\produce@RRAP{*#1\href{mailto:#2}{#2}}}\frontmatter@RRAPformat}
  {}{}
}%
\makeatother
\begin{document}

\preprint{AIP/123-QED}

\title{Combining physics-based and data-driven predictions for quantitatively accurate models that extrapolate well; with application to DIII-D, AUG, and ITER tokamak fusion reactors}
\author{J. Abbate}
\affiliation{PPPL}
\author{E. Fable}
\affiliation{Max Planck Institut für Plasmaphysik}
\author{G. Tardini}
\affiliation{Max Planck Institut für Plasmaphysik}
\author{R. Fischer}
\affiliation{Max Planck Institut für Plasmaphysik}
\author{E. Kolemen}
\affiliation{PPPL}
\author{ASDEX Upgrade Team}
\affiliation{Max Planck Institut für Plasmaphysik}

\date{\today}

\begin{abstract}

Methodologies for combining the accuracy of data-driven models with extrapolability of physics-based models are described and tested, for the task of building transport models of tokamak fusion reactors that extrapolate well to new operational regimes. Information from multiple physics simulations (the ASTRA transport code with gyro-Bohm and TGLF estimates for turbulent diffusion) as well as multiple distinct experiments ({DIII-D} and AUG tokamaks) are considered. Applications of the methodology to the task of commissioning and controlling a new reactor such as ITER are discussed.

\end{abstract}

\maketitle

\section{Introduction}

Commissioning and operating reactor-class fusion experiments like ITER~\cite{claessens_michael_iter_2020} will offer unique challenges beyond what the plasma physics community has so far experienced. As in previous experiments, operators must decide how to set each actuator waveform over time in order to attempt to achieve a desired plasma state. Unlike in previous experiments, however, there will be fewer discharges from which to learn due to the higher cost per discharge; decisions will need to be made more quickly due to the tighter experimental timelines; greater extrapolation to novel physics regimes will be necessary; and all of this must be done with greater reliability due to increased nuclear licensing requirements and risk of machine damage. 

Like any experimental campaign, ITER will have a staged set of operations during which plasma parameters are increased. The progression through phases, primarily focused on scaling from low to high plasma current, will be as gradual as allowed by the tight timeline and limited number of discharges~\cite{iter_iter_2018}. A key question is how to use the massive amount of data acquired from each discharge to modify and enhance the decisions for the next discharge in order to give the greatest confidence of achieving the target plasma state without damaging the device. A related question is how to build algorithms that quickly make decisions in realtime, without human input, about how to modify actuators given the trajectory of the plasma state all the way up to the present timestep. The information should be considered alongside all other knowledge available at the time of the decision. This includes all experimental data from historic devices like the {DIII-D}~\cite{luxon_big_1985} and {AUG}~\cite{keilhacker_asdex_1985} tokamaks. It also includes all sufficiently fast physics calculations and simulations, which contain e.g. assumptions about local conservation of energy, momentum, and particles. Consider that ``sufficiently fast" may require computability within months for inter-phase planning, days for intra-phase planning, minutes for inter-discharge planning, and seconds for (realtime) intra-discharge planning. 

One useful tool in decision-making (especially when done algorithmically rather than by human debate) is forecasting: predicting the evolution of the plasma state given the present state and a future trajectory of proposed actuators. With such a model, uncertainty quantification and scenarios analysis can be done between discharges to aid human decision-making and tuning of controllers, and model-predictive control can be readily deployed within discharges to modify actuator trajectories while respecting actuator and state constraints. One area where this is especially important is kinetic profile control, where physics is high-dimensional, nonlinear, and crucial to high-level metrics for ITER's operational phases like performance, stability, and heat flux management. Kinetic profile control involves tailoring the full profiles of density (electron and all ions including impurities), momentum, temperature, and plasma current. This type of multi-input multi-output (MIMO) control is difficult and has not been tackled very much on previous devices, but will be especially important for ITER's steady-state scenario mission \cite{snipes_physics_2014, humphreys_novel_2015}. 

Prior work in kinetic profile control relies on the assumption of the validity of transport equations, along with various empirical approximations for the coefficients of the simulation \cite{ou_towards_2007, felici_non-linear_2012, morishita_first_2024}. Though these ad hoc adjustments can be made before the discharge and additionally as it progresses, if the underlying assumptions of the model are largely wrong and merely adapted to look correct, then the physics models have no more extrapolability than an empirical model. What is more, predicting all components of the plasma state has so far been elusive, e.g. due to difficulty predicting densities due to lack of information about the wall conditions~\cite{abbate_large-database_2024}. For example, it has been recognized by the STEP reactor planning team that no kinetic profile evolution model is sufficiently validated at present to be expected to predict a reactor \cite{mitchell_scenario_2023}.

As a role-play of the extrapolation that will be used during ITER commissioning for decision-making, this work demonstrates and validates various such physics models for evolving kinetic profiles for various subsets of {DIII-D} data. Following up on \cite{abbate_data-driven_2021}, it also demonstrates and compares an empirical analogue (based on a neural network) trained on experimental data only. This validation is done not only for accuracy within-distribution, where empirical models could surely do better, but for extrapolating to nearby regimes, as will be necessary for the staged approach to ITER operations. This validation is also done over a large database of randomly selected shots to avoid bias. It is demonstrated that when using a meta-learned ensemble of both data-driven models and physics-based simulators, performance in predicting out-of-distribution plasma profile evolution is better than either one individually. Various alternative methodologies for augmenting machine learning predictions with simulations and data from similar experiments ({AUG}) are also considered, though demonstrated to yield no significant performance enhancements. 

In Section~\ref{sec:methodology}, the mechanisms used in this paper for validating the performance of predictive kinetic profile models during staged commissioning and operation of a device is described. In Section~\ref{sec:physics}, the ASTRA transport solver \cite{pereverzev_astra_1991} setup used as the physics-based simulator is described. In Section~\ref{sec:machinelearning}, the method for building an analogous empirical model with a neural network is presented. In Section~\ref{sec:results}, mechanisms and corresponding results are presented for fusing information from the various physics and empirical models.

\section{Methodology}
\label{sec:methodology}

\subsection{Validation}
For validation and comparison of plasma evolution models to experimental reality, the $\sigma$ metric outlined by the ITER working group in 1999 \cite{iter_physics_expert_group_on_confinement_and_transport_chapter_1999} will be employed, which measures percentage discrepancy in a plasma profile $X$ over radial coordinates $j$ as
\begin{equation}
    \sigma =  100\% * \frac{ \sqrt{ \frac{1}{N} \sum_{j=1}^N \left(X_j-X_{\text{truth},j}\right)^2 } }{\sqrt{ \frac{1}{N} \sum_{j=1}^N X_{\text{truth},j}^2 } } 
    \label{eq:sigma}
\end{equation}

In this work, predictions are made exactly 300ms from an initial time. The $\sigma$ metric can be considered over time, i.e. as a function of the length of time $\Delta t$ into the future of the prediction. $\sigma$ can also be considered as an average value over the full prediction trajectory, where in this work all timesteps are weighted equally across the 300ms window. 300ms was chosen as a window relevant for kinetic profile control: the energy confinement time at {DIII-D} and AUG is order 50-100ms~\cite{abbate_data-driven_2021}. 

Note that machine learning models in this work are not trained on exactly this metric, and instead on total mean-squared error of normalized profiles. For simplicity (in contrast to prior work) the normalization is simple division by a number that makes the profiles order unity (for example, a factor of 100 is used for rotation as measured in $\frac{krad}{s}$, because the quantity in {DIII-D} and {AUG} tends to be order of tens to lower hundreds).

\subsection{The plasma state and the controlled actuators}
\label{sec:state_and_actuators}
The relevant data for control generally consists of the plasma state and plasma actuators. Relevant plasma state parameters (both the parameters being controlled and relevant ``hidden or ``context" variables necessary for evolving those parameters) are given up to the initial time and then must be autoregressively predicted by the model through the end of the prediction window. The controlled actuators are given at all times during prediction. In the specific case of kinetic profile modeling, which is the task tackled by transport models and the analogous empirical machine learning models developed in this study, the plasma state is considered as 6 one-dimensional profiles:
\begin{enumerate}
    \item Electron temperature, $T_e$
    \item Ion temperature, $T_i$ (assumed to be the same for all ion species in this study)
    \item Electron density, $n_e$
    \item Effective charge $Z_{eff}=\sum_s n_s Z_s^2$ for $n_s$ the ion density of ion species $s$ and $Z_s$ the charge of the ion species
    \item Safety factor $q$
    \item Plasma rotation frequency $\Omega$
\end{enumerate}

In both {AUG} and {DIII-D}, $T_e$ and $n_e$ are primarily measured and fitted from Thomson scattering data, while $T_i$ and $\Omega$ from charge exchange recombination data. For {DIII-D}, $Z_{eff}$ is calculated by assuming that Carbon is the only impurity present in the plasma, that all of the Carbon is fully stripped, and that the plasma is quasineutral. $Z_{eff}$ is then readily calculated using $n_e$ (Thomson) and $n_C$ (Charge Exchange Recombination) data. Meanwhile, safety factor $q$ is reconstructed by the standard EFIT01 workflow~\cite{lao_mhd_2005}, which constrains the Grad-Shafranov equation with external magnetics measurements only. As discussed in \cite{abbate_data-driven_2021}, internal constraints (the so-called EFIT02 workflow) were avoided due to the empirical observation that neural networks trained on EFIT01 data perform better. For {AUG}, the $Z_{eff}$ profile is instead estimated based on bolometric data; and $q$ is additionally constrained by core measurements from Motional Stark Effect (for current density) and plasma density and temperature (for pressure). More details about the standard ``integrated data" workflows used to generate the AUG profiles can be found in~\cite{fischer_integrated_2010,rathgeber_estimation_2010,fischer_coupling_2016,fischer_estimation_2020}.

The actuators that are considered for directing the evolution of the above plasma state are as follows:
\begin{enumerate}
    \item Voltage, power, and duty cycle of each neutral beam (NB) injector along with the geometry, or in reduced form total NB power $P_{NBI}$ and torque $\tau_{NBI}$
    \item Frequency, power, and injection geometry of each electron-cyclotron heating (ECH) gyrotron, or in reduced form total ECH power $P_{ECH}$
    \item Toroidal magnetic field $B_t$
    \item Total plasma current $I_p$
    \item Plasma last-closed flux surface (LCFS) boundary contour points $(R,Z)$, which can be further reduced to the 0\textsuperscript{th} through 3\textsuperscript{rd} (excluding Shafranov shift, which is 0 for the LCFS)``moments" of the shape: $R$ the major radius, $a$ the minor radius, $\kappa$ the elongation, $\delta_l$ the lower triangularity, and $\delta_u$ the upper triangularity
    \item Volume- or line-averaged density $<n_e>$, and/or total puffed Deuterium from all valves ``D Gas".
\end{enumerate}

It is important to consider that many of these ``actuators" for kinetic profiles really would be state variables for a lower-level controller that does not always accurately achieve the desired setpoint. For example, the last-closed flux surface boundary is (on both current devices and as presently envisaged for ITER) controlled by feedback on spatial setpoints via the numerous poloidal field coils; plasma current is fed back on a Rogowski coil measurement via the voltage applied to the central solenoid; etc. The least controllable among the actuators listed is perhaps the volume-averaged density, which is usually actuated with a proportional controller on the gas-puffing valves. For this and all actuators, one would ideally use the target state value for the controller rather than the measured value. However, when mapping between machines with different control systems and different diagnostic geometries and calibrations, this becomes difficult. What is more, the density is only sometimes controlled: in some discharges the gas valves are instead set with feedforward values, or density feedback is employed but with some feedforward gas valves fixed. And either way, especially in a carbon-walled device like {DIII-D} where outgassing can outpace pumping, there is often no controllability for decreasing density, only for increasing it. For these reasons, the models considered in this paper use total puffed Deuterium Gas but not target density as an actuator. 

The evolution of kinetic profiles in the absence of macro-instabilities takes place on timescales of roughly 100ms on devices like {DIII-D} and {AUG}~\cite{abbate_data-driven_2021}. Necessary diagnostics for profiles come in as infrequently as every 10ms. So for validating models (and in the case of empirical models, also for training them) data is considered every 20ms. The timescale for the predictions can be as long as desired for validating the empirical models, but due to practical limitations of current primarily-physics-based models (as will be discussed in Section~\ref{sec:physics}) a specific prediction window of 300ms is used when comparing simulations alongside empirical models.

\subsection{Validation datasets}
All empirical models are built and tuned based on randomly selected experimental data from timesteps where all data is available. However, a limitation of these models is that they are optimized for a given scenario and may not perform well in unseen discharges. This work seeks to simulate validating models on unseen discharges to understand their accuracy for reactor-grade startup in predicting ``in-distribution" (corresponding to rerunning a discharge) vs ``out-of-distribution" (corresponding to planning and controlling future discharges in unseen regions of phase space). Though ITER and other reactor-class devices will commission and gradually increase a variety of actuators, for simplicity this paper focuses on the progression of just increasing plasma current $I_p$, with three regions of phase space considered throughout this paper, with the following training dataset (which excludes discharges ending in 0 and 5 for test and validation sets):
\begin{enumerate}
    \item $\mathbf{I_p<0.9MA}$: 820,853 timesteps from 13,572 discharges
    \item $\mathbf{I_p<1.2MA}$: 2,049,309 timesteps from 17,243 discharges
\end{enumerate}

Models trained using $I_p$<0.9MA samples can be tested ``out-of-distribution" on other cases with $I_p$ higher, such as cases between 1.0MA and 1.2MA. Similarly, models trained with $I_p$<1.2MA can be trained on cases with $I_p$ greater than 1.3MA. 
A subset of 300ms predictions chunks is used for these validations:
\begin{enumerate}
    \item $\mathbf{1.0MA<I_p<1.2MA}$: 18,150 timesteps from 196 discharges
    \item $\mathbf{1.3MA<I_p}$: 25,800 timesteps from 309 discharges
\end{enumerate}

\begin{figure}
    \centering
    \includegraphics[width=0.5\textwidth]{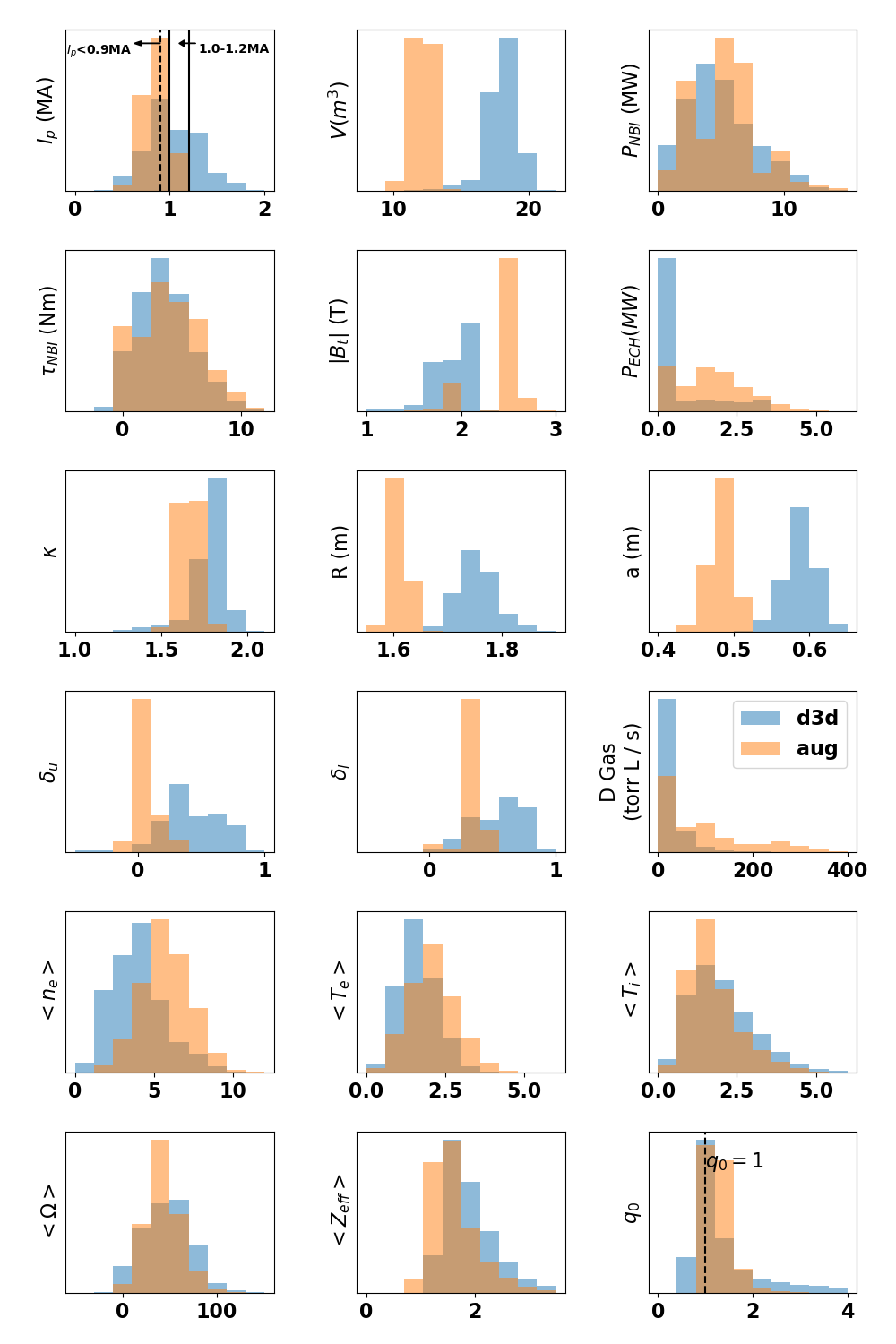}
    \caption{Histogram comparing the distribution of values of both actuators and profiles used in this study for {DIII-D} (blue) vs AUG (orange). Note that the primary statistical difference is the size of the devices, the toroidal field, the gas and density, and the shaping. For reference, the $q_0=1$ surface and the $I_p<0.9MA$ and $1.0MA<I_p<1.2MA$ regions are shown on the corresponding plots.}
    \label{fig:aug_d3d_histogram}
\end{figure}

For training models using AUG data, 265,632 timesteps from 1041 unique shots from AUG cases up to about $I_p$>1.2MA are added to the DIII-D $I_p$<0.9MA dataset. For reference, the distribution of {DIII-D} vs {AUG} data used in this study are shown in Figure~\ref{fig:aug_d3d_histogram}. Note the higher Deuterium gas puffing in AUG, likely due to the outgassing of the porous carbon wall on {DIII-D}; the higher toroidal field magnitude $B_t$ of {AUG}; and the larger size of {DIII-D}. Shaping on {DIII-D}, including elongation $\kappa$ and upper/lower triangularities $\delta$, is slightly more extreme (noncircular). $T_e$ is slightly higher at {AUG}, largely due to the higher ECH power. Density in AUG is also slightly higher on average. Additionally, the distribution of core safety factor $q_0$ is fairly different, though this is likely due to differences in differences in the reconstructions used for the different devices. The $I_p<0.9MA$ and 1.0MA<$I_p$<1.2MA regions used for some of the validation in this work are shown for reference. 

\section{Physics models}
\label{sec:physics}
For full-discharge kinetic profile evolution, the state-of-the-art simulators are based on transport-like coupled partial differential equations (PDEs). The general idea is that the one-dimensional profiles (for any profile $X$) are assumed to evolve ($\sim \frac{\partial X}{\partial t}$) with a combination of a source term, diffusion ($\sim - \nabla^2 X$), and pinch ($\sim - \nabla X$). These equations collectively account for the conservation of particles, momentum, energy, and magnetic flux in the system. In tandem with these PDEs, the Grad-Shafranov equation is also solved with the last-closed-flux-surface boundary as input (or alternatively with coil currents for a free boundary solution, though this will not be considered in this work). An open-source module was developed for this work and is available in OMFIT~\cite{meneghini_integrated_2015} to prepare and launch the implementations of all of these options via the ASTRA~\cite{pereverzev_astra_1991} transport modeling framework. For simplicity, only predictions of electron temperature $T_e$, ion temperature $T_i$, and ion rotation $\Omega$ are considered.

Heat and momentum deposited by neutral beams is estimated with the RABBIT~\cite{weiland_rabbit_2018} code. Heat deposited by electron cyclotron heat is calculated with the TORBEAM~\cite{poli_torbeam_2001,poli_torbeam_2018} code. Internally, ASTRA also calculates electron-ion heat exchange, ohmic heating, and radiation heat loss.

For diffusion, two distinct physics models are considered: 

First, the simpler ``GyroBohm" scaling
\begin{equation}
    \chi_{GB} \sim \frac{T_e^\frac{1}{2}}{B_T^2}\nabla T_e
\end{equation}
has often been used in an empirically determined linear combination with a nonlocal ``Bohm" term, $\chi_B \sim \frac{q^2}{n_e B_T} \frac{dp_e}{dr} \frac{T_e(\rho=0.8)-T_e(\rho=1.0)}{T_e(\rho=1.0)}$, for the so-called Bohm gyroBohm model~\cite{erba_validation_1998}. However, these parameters need to be finely tuned to the specific regime of interest, and the nonlocal term is not considered in this work. 

Second, TGLF~\cite{kinsey_first_2008} is meant to be a fully-physics-based model for determining the expected outward flux of heat, particles, and momentum due to turbulence. It uses a quasilinear approximation such that calculations take order of minutes, in contrast to the full gyrokinetics simulations taking order of days or months. It uses empirical fits to gyrokinetics rather than to experiments for determining nonlinear saturation thresholds. For simplicity, in this work a neural network surrogate~\cite{meneghini_self-consistent_2017} for TGLF saturation rule zero is considered. The model is local, taking as inputs 24 parameters such as $\nabla T$, $\frac{dq}{d\rho}$, $\beta$; and returning the local expected flux. Developed to predict tokamaks seen thus far when used in transport solvers, the model is primarily built to account for Ion Temperature Gradient (ITG), Electron Temperature Gradient (ETG), and Kinetic Ballooning Mode (KBM) turbulence.

One can use the same transport codes used for predicting profiles (ASTRA in this case) to ``interpret" the diffusion coefficients that would yield the experimentally measured profile evolution, by assuming the estimates for heat and momentum deposition are roughly correct and that only the diffusion coefficient is unknown. The profiles and actuators for timesteps up to the moment at which a prediction is needed can be used to estimate the initial diffusion coefficients as
\begin{subequations}
\label{eq:diffusion_coefficients}
    \begin{equation}
       \chi_\text{ion heat,Expt}=\frac{1}{n_i\frac{dT_i}{d\rho}} \frac{\int_0^\rho S_\text{ion heat} dV-\frac{d\int_0^\rho(n_iT_i)dV}{dt}}{\frac{dV}{d\rho} <\nabla \rho^2>} 
    \end{equation}
    \begin{equation}
       \chi_\text{electron heat,Expt}=\frac{1}{n_e\frac{dT_e}{d\rho}} \frac{\int_0^\rho S_\text{electron heat} dV-\frac{d\int_0^\rho(n_e T_e)dV}{dt}}{\frac{dV}{d\rho} <\nabla \rho^2>}. 
    \end{equation}
\end{subequations}
These initial diffusion coefficients can then be adjusted to set the model's diffusion coefficient across time $\chi(t)$ to be the experimentally determined initial $\chi_{Expt}(t=0)$ scaled with the physics estimate $\chi_{Phys}$:
\begin{equation}
    \chi = \chi_\text{Expt}(t=0) \frac{\chi_{Phys}(t)}{\chi_{Phys}(t=0)}
\end{equation} 

For all predictions used in this work, $\Omega$ evolution employs momentum diffusion fixed at its initial value without a physics adjustment. For $T_e$ and $T_i$ prediction, TGLF-nn is employed as-is (``tglf-nn"), and is also considered within the physics-adjusted fixed-diffusion framework (``fixed+tglfnn"). Finally, because an empirical estimate for the initial condition is needed for the gyroBohm model, the gyroBohm model is employed only within the physics-adjusted fixed-diffusion framework (``fixed+gyrobohm").

\section{Machine learning models}
\label{sec:machinelearning}

\begin{figure}
    \centering
    \includegraphics[width=0.5\textwidth]{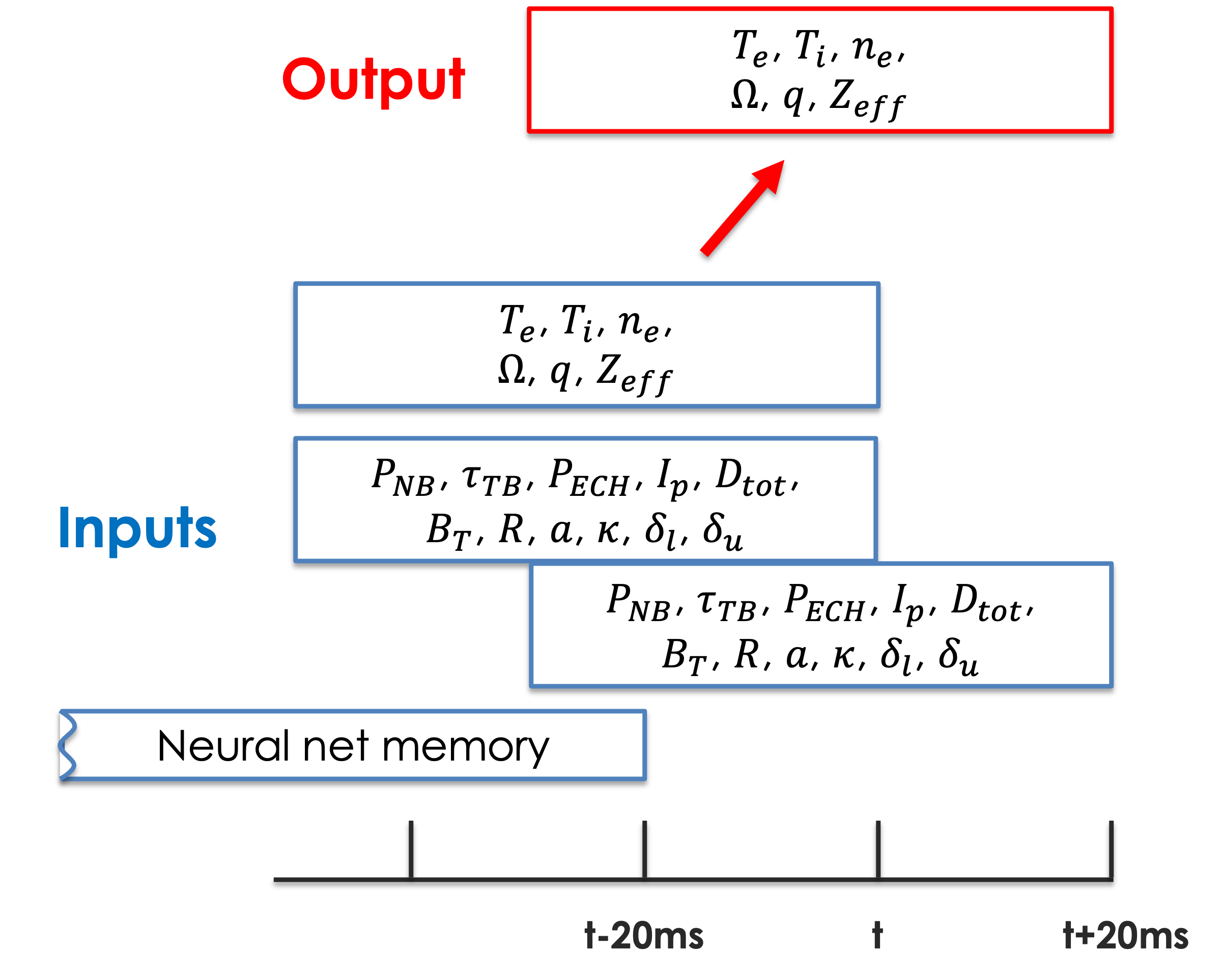}
    \caption{Machine learning model inputs and outputs. Note that all values considered are discretized to steps $\Delta t_\text{step}=20ms$, and with the box-car average of all measurements within the timestep and a window of size $\Delta t_\text{smooth}=50ms$ prior (i.e. causal smoothing). Specifically, profiles at the present timestep t are input as smoothed values between t and t-$\Delta t_\text{smooth}$. Actuator values at timestep t are input as both present and future information, i.e. values at t+$\Delta t_\text{step}$ (averaged between t+$\Delta t_\text{step}$ and t+$\Delta t_\text{step}$-$\Delta t_\text{smooth}$) and also at t (averaged between t and t-$\Delta t_\text{smooth}$). Finally, the LSTM's own recurrent memory is input, which considers information prior to t-$\Delta t_\text{step}$. The model is optimized to output an estimate for the profiles at t+$\Delta t_\text{step}$ (averaged between t+$\Delta t_\text{step}$ and t+$\Delta t_\text{step}$-$\Delta t_\text{smooth}$). To predict multiple timesteps, autoregression (i.e. giving the output profile from the previous timestep as input for the next timestep) is employed. This is analogous to the inputs, outputs, and autoregression process for PDE-based simulators.}
    \label{fig:input_output}
\end{figure}

A neural network for discretely predicting the change in plasma profiles was described and validated~\cite{abbate_data-driven_2021}, and used for simple model-predictive control on {DIII-D}~\cite{abbate_general_2023}. A followup model used autoregressive rollout (i.e. taking small steps but feeding back on its own predictions so that predictions can be made arbitrary steps into the future), uncertainty predictions (i.e. outputting a mean and a standard deviation), and ensembling (i.e. training many models with the same parameters and using the average of the ensemble as the prediction, with standard deviation for reference)~\cite{char_offline_2023}. In the latter model, the inputs to the network at each timestep are the profiles, the actuators at the present timestep, and the actuators at the next timestep (Figure~\ref{fig:input_output}). This work largely follows that methodology, but using more profiles and actuators (listed in Section~\ref{sec:state_and_actuators}) and employing a more advanced training mechanism to optimize the model for longer time-horizon predictions (described below). The model architecture follows closely previous approaches, in this instantiation with an \textbf{encoder} consisting of two fully connected layers (total of 1,224,000 parameters), followed by an \textbf{RNN} (a standard LSTM~\cite{hochreiter_long_1997}, with 440,800 parameters), followed by a \textbf{decoder} of three fully connected layers (total of 1,300,198 parameters). In total, then, the model has 2,964,998 parameters; this is the same order of magnitude as the number of timesteps used to train the models (see Section~\ref{sec:methodology}). The model hyperparameters are held fixed for all demonstrations in this paper. PyTorch\cite{paszke_pytorch_2019} is leveraged to train the networks.

\begin{figure}
    \centering
    \includegraphics[width=0.5\textwidth]{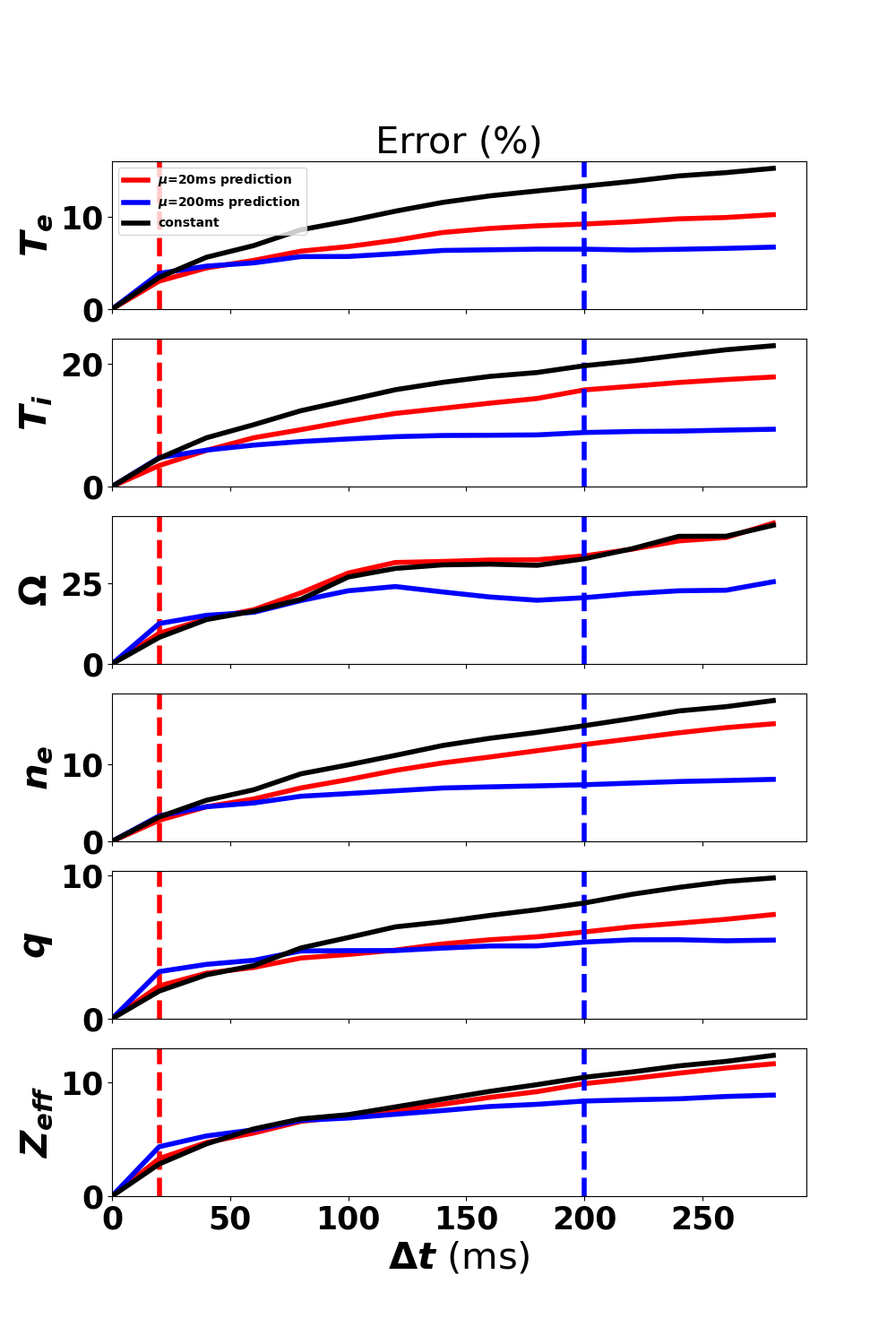}
    \caption{Via curriculum learning, the model can be tuned to arbitrary number of timesteps. Here are shown models optimized for a single timestep (20ms, in red), which has lower error $\sigma$ at 20ms (red dashed line), and 10 timesteps (200ms, in blue), which has lower error at later times (blue dashed line). For reference, error for predictions where profiles are assumed to stay constant at the initial values is also plotted, in black.}
    \label{fig:curriculum}
\end{figure}

The simplest mechanism for training an autoregressive model is teacher forcing, where during training the model is optimized for predicting only the next timestep, but during testing the model must use its own output to predict later timesteps. In curriculum learning \cite{bengio_scheduled_2015}, training begins exclusively with teacher forcing  but then gradually during training the model must use its own outputs autoregressively for longer and longer timesteps. Specifically, at each timestep during training it is randomly decided with a probability $1/\mu$ whether to use the true present timestep or use the prediction from the previous step. The number of steps taken autoregressively before the experimental value is once again ``forced" therefore follows an exponential distribution where $\mu$ is the mean number of steps. The model can therefore be optimized to do best at a user-specified number of timesteps. In Figure~\ref{fig:curriculum}, $\sigma$ error (Equation~\ref{eq:sigma}) as a function of the number of timesteps predicted ($\Delta t$) is shown (for predictions on an in-distribution test set) for the same model optimized for 20ms predictions (red) and 200ms predictions (blue). For reference, the baseline error for ``constant" predictions (assuming no change in profiles from the initial time) is shown in black. The error for the 20ms model is lower at 20ms, but is higher for later times. In this work, the model is optimized to predict with a mean of 200ms ahead. This was chosen to be a bit less than the 300ms prediction window used in the paper, since average accuracy through the entire window is what is desired (and what is validated against in Section~\ref{sec:results}).

\begin{figure}
    \centering
    \includegraphics[width=0.5\textwidth]{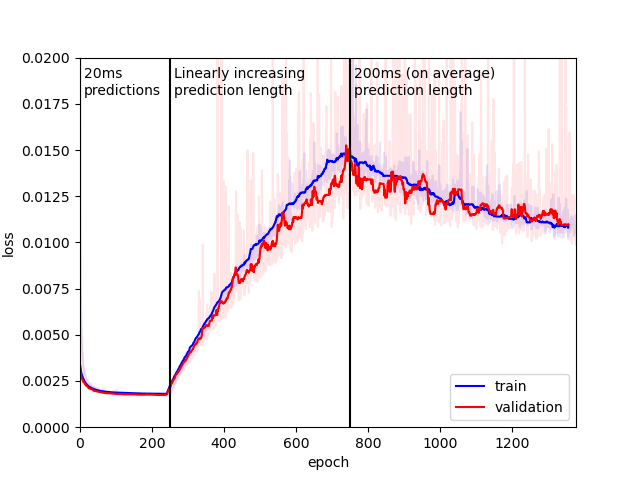}
    \caption{Typical loss curve for curriculum learning. For the first 250 epochs teacher forcing is used, and the model need only predict one timestep (20ms) ahead. The model then begins randomly using its own output autoregressively as input for predicting the next timestep, with an increasing probability of autoregression (i.e. average number of steps per autoregressive rollout) between epoch 250 and 750 from 20ms up to 200ms. Between epoch 750 and the end of training a probability corresponding to an average number of steps of 200ms is used.}
    \label{fig:loss}
\end{figure}

During training, we start by teacher forcing for 250 epochs, then linearly progress from a mean of $\mu=1$ step (20ms) to a mean of $\mu=10$ steps (200ms) for 500 epochs, then allow the model to learn with the average of $\mu=10$ steps fixed for another 750 epochs. The model with the lowest validation loss after epoch 750 is used (i.e. ``early stopping"). An illustration of a typical loss curve is shown in Figure~\ref{fig:loss}, demonstrating the task getting gradually harder as the model must make longer time-horizon predictions without teacher forcing between 250 and 750 epochs until it is able to optimize with the task fixed afterwards. Note that the validation set in this case is all discharges ending in 5, with all discharges ending in 0 reserved for testing to give the final plots in Section~\ref{sec:results}. 

\section{Combining information from simulations and other machines to improve extrapolability}
\label{sec:results}

Ultimately, for control and decision-making during commissioning of a device like ITER, one needs models that are not only accurate but extrapolate well and describe causality with actuators. Empirical models are expected to be most accurate for in-distribution cases. However, they are not guaranteed to maintain causality and are only able to learn the cases they see. More generally, they are not expected to extrapolate to near- and far-distribution cases.

Abstractly speaking, humans have collectively been able to develop tokamak physics models by making hypotheses that can be tested with experiments. The philosophy of many transport models is that one should keep developing theories until results match experimental data on tokamaks specifically, which can be subject to the same overfitting as machine learning due to insufficient quantities of tokamak data. Nonetheless, human intuition from many other unrelated experiments (e.g. watching planetary motion to validate basic force equations which have been seen to extrapolate well to many domains) endows physics-based models with a distinct advantage. This section describes techniques for both adding more experimental data (by supplementing with AUG discharges) and adding information from simulations to attempt to achieve better extrapolability to near- and far-distribution cases. 

The following are the mechanisms for extrapolation that will be demonstrated in this section:
\begin{enumerate}
    \item \textbf{Augment dataset} with data from other experiments, in this case AUG. One may want to find better ways to map signals from one device to another, which can be considered an exercise in feature engineering. Though feature engineering should not be necessary for large enough models and datasets (it was primarily an old technique needed for shallow learning algorithms), it may help in the case of making models better extrapolate. Physicists particularly like non-dimensionalization of parameters, which is effectively employed by the widely acclaimed confinement scaling estimates (e.g. H98) \cite{hugill_empirical_1978, iter_physics_expert_group_on_confinement_and_transport_chapter_1999} and will be discussed in more detail in this section.
    \item \textbf{Add simulation context as machine learning inputs}, such as interpreted and predicted values from transport simulations. This is similar to work done in Refs.~\onlinecite{karpatne_theory-guided_2017, baier_hybrid_2021} to add simulation information as input and to predict differences between simulation predictions and reality instead of the reality itself. Within plasma physics, this tactic has also been used for Inertial Confinement Fusion. \cite{humbird_cognitive_2021} 
    \item \textbf{Transfer learn} by training on experimental data and tuning on simulation data as if it were experimental data. Similar work has been attempted on NIF in Ref.~\onlinecite{humbird_transfer_2020}, though with the opposite mindset of training on simulation data and tuning on experimental data. 
    \item \textbf{Stacked generalization}, or creating a meta-learned model to optimally extrapolate from an ensemble of predictors (both machine learning and simulations in this case). The key advantage of this approach is that the cost function of the meta-learning algorithm specifically optimizes for the task of predicting to unseen regimes from models built within seen regimes.
\end{enumerate}

It should be noted that another potential technique for adding knowledge to empirical models is providing extra constraints in the cost function, most popularly energy conservation as in \cite{jia_physics-guided_2021}. However, for highly-complicated fusion grade plasmas where not everything is measured, the authors cannot think of any good global constraints that could aid predictions. The closest approximation to this is perhaps the particle, momentum, and energy balance that the transport models attempt to approximate.

In this section, various techniques drawing from these ideas are described, and results presented. Specifically, as described in Section~\ref{sec:methodology}, the $\sigma$ metric averaged over the 300ms prediction windows will be shown for various cases predicting in-distribution and far-distribution. Two cases will be shown: in blue will be the performance on the full distribution, and in red will be the performance on the subset of trajectories with the difference in injected power $\Delta P_{inj}$ greater than 500kW between the average of the first 100ms and the average of the last 100ms. This selects more dynamical cases, and gives some sense of the ability of the model to predict actuator responses (causality) instead of simply learning correlations. In all plots, dashed horizontal lines are plotted as the null hypothesis performance of simply predicting that profiles do not change from the initial condition. Horizontal black lines around the bar plots indicate the 25th and 75th percentiles of the distribution of $\sigma$ error. 

\subsection{Add AUG data (with normalization)}

First, machine learning models trained only on {DIII-D} data are compared to models trained with {DIII-D} in addition to AUG data, and the predictions are validated on a near-distribution {DIII-D} dataset. As shown in Figure~\ref{fig:aug_d3d_histogram}, {DIII-D} has a Carbon wall, a lower maximum toroidal magnetic field, and a larger size. With these slight differences in mind, the non-dimensionalization mentioned above is considered to aid in the generalization of the machine learning model. 

Non-dimensionalization of basic plasma physics equations theoretically allows ``wind-tunnel" scaling between different regimes: for a simplified device the core-transport physics is theoretically the same given the same $\rho^*$, $\beta$, $\nu^*$, $\epsilon$, $q$, and $T_e/T_i$.~\cite{kadomtsev_tokamaks_1975} However, it turns out radiation (atomic physics), wall interaction, 3D field effects, etc. play a significant role which increases the number of self-similar parameters that need to be matched (see Ref.~\onlinecite{petty_sizing_2008} for an overview). What is more, difficulties arise in creating a uniformly invertible mapping due especially to the fact that profile values near the edge are close to zero. An empirical but widely acclaimed (especially for control room operation) normalization is the Greenwald fraction~\cite{greenwald_density_2002} for density, $n_{GW} \sim \frac{I_p}{\pi a^2}$ (a dimensional quantity, with $I_p$ in MA, $a$ in m, and density in $10^{20}m^{-3}$). Although this is an empirical scaling still seeking theoretical understanding, it is used frequently in tokamak control rooms for its use in predicting a density limit around and past which performance degrades and disruptions become more likely. Theoretically, with enough data from different devices, the machine learning model could learn this scaling on its own; but with just one device extrapolating to another it is hypothesized scaling by this quantity could help. 

An additional normalization considers the expected mapping between actuators and profiles absent complicated physics: power should scale like thermal energy (rather than temperature itself), and torque should scale like momentum (rather than rotation itself). In effect, this can be considered as simple feature engineering, but with an eye toward extrapolation. 

In this work, we ultimately chose to consider the following normalization:
\begin{align}
    n_e &\rightarrow \frac{n_e}{n_{GW}} \\
    \Omega &\rightarrow \int n_e dV R^2 \Omega \\
    P_{inj} &\rightarrow \frac{P_{inj}}{V}
\end{align}
where $V$ is plasma volume, and $\int n_e dV$ is a rough approximation for the plasma mass. This approximation is based on the assumption that mass scales like ion charge (which is true for fully-stripped species where the ratio of neutrons and protons is similar) and that quasineutrality holds. Additionally, to approximate a volume integral, we consider that the volume of each shell scales like $\rho_n$ so that the total integral scales like $(\vec{n_e} \cdot \vec{\rho_n}) V$. In particular, the core density is weighted less than the edge density with linear proportionality. {DIII-D} primarily contains fully stripped Carbon impurities, while AUG contains Tungsten which has a variety of charge states.

\begin{figure}
    \centering
    \includegraphics[width=0.5\textwidth]{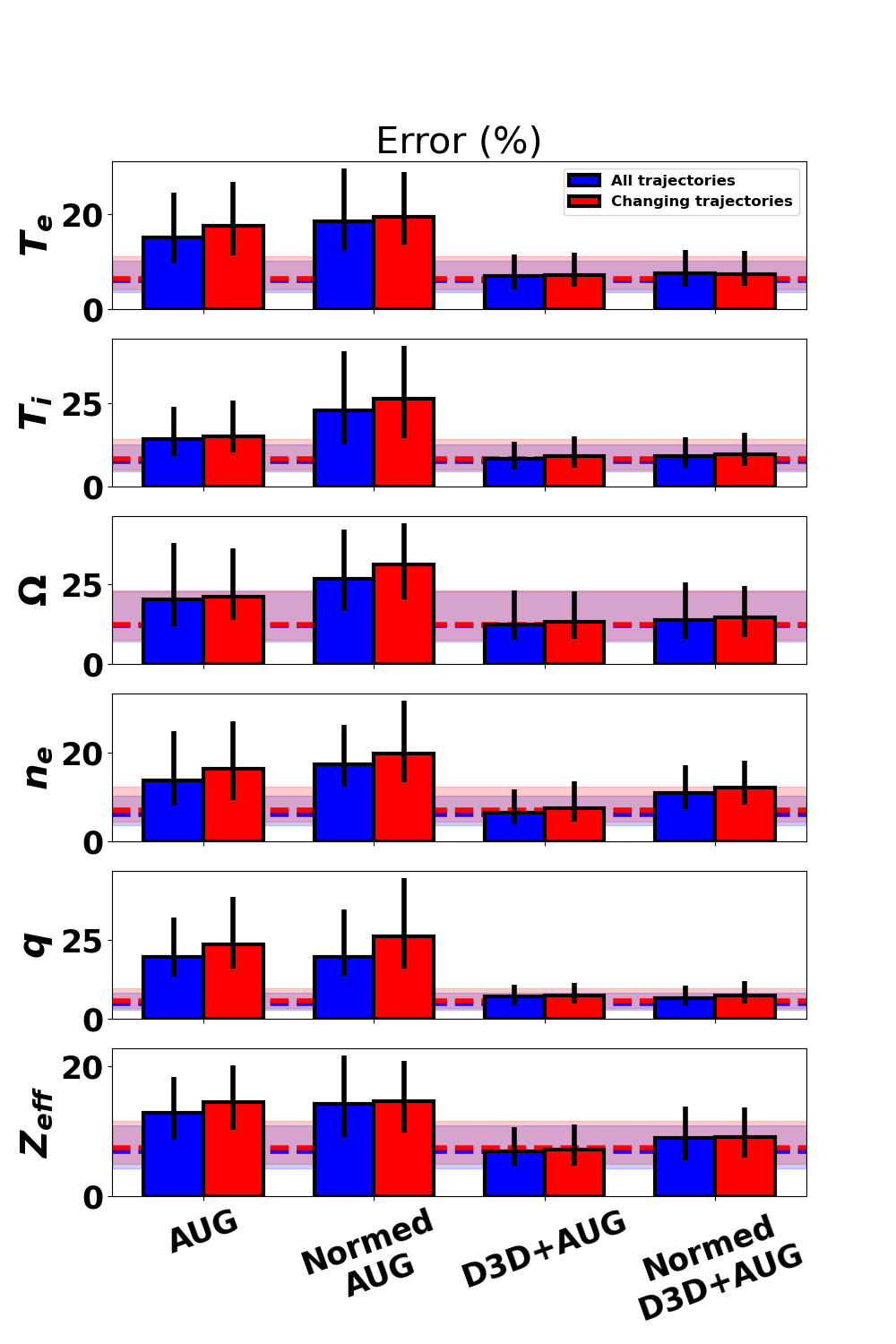}
    \caption{Comparing $\sigma$ error (Equation~\ref{eq:sigma}) with AUG database with and without normalization, along with bleeding AUG data into {DIII-D} data.}
    \label{fig:augSigmaComparison}
\end{figure}

Ultimately, whether with or without the normalization, adding AUG data makes no significant improvement in near-distribution profiles. As shown in Figure~\ref{fig:augSigmaComparison} where $\sigma$ error averaged over 300ms predictions (as described in Section~\ref{sec:methodology}) is considered for predicting {DIII-D} cases for $1.0MA<I_p<1.2MA$. One machine learning model is trained with {DIII-D} data only for $I_p<0.9MA$ to test near-distribution predictions, and another for $I_p<1.2MA$ for in-distribution comparison. Another model trained on both {DIII-D} $I_p<0.9MA$ alongside {AUG} cases (which range up to about $1.2MA$) is considered, both with and without normalization, and in both cases the performance is no better than when trained on the {DIII-D} $I_p<0.9MA$ cases alone. For reference, the performance of the model when extrapolating from AUG to {DIII-D} directly (with and without normalization) is shown as well. For such machine-to-machine extrapolation, the performance is generally not better than predicting that profiles never change.

These initial results suggest the difficulty of predicting very far out of distribution (all the way to a brand new machine), and demonstrate the reason reactors like ITER plan to commission as gradually as possible, and use their own data to learn rather than rely on much previous experimental data. 

\subsection{Transfer learning}
Another option is to begin with a machine learning model trained on experimental data, then tune on simulation outputs as if they are experimental truth. In other words, the simulation output is used in place of the experimental truth in the cost function during training. Once again, the initial machine learning model is trained in-distribution ($I_p$<0.9MA). The model is then tuned with example simulations that include data from the near-distribution set ($I_p$<1.2MA). The tuned model is then validated for the near-distribution set, 1.0MA<$I_p$<1.2MA. 

\begin{figure}
    \centering
    \includegraphics[width=0.3\textwidth]{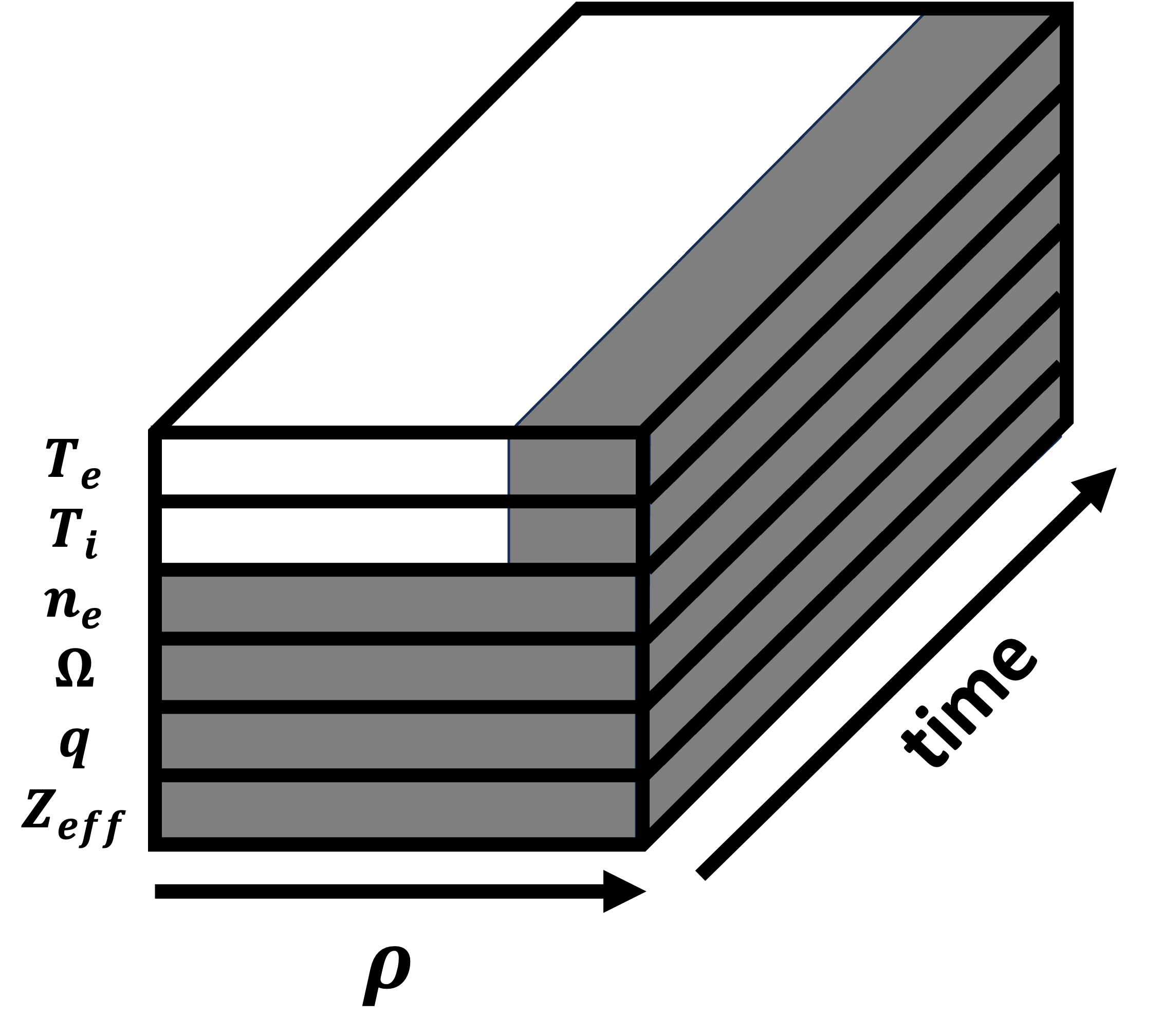}
    \caption{Mask used to tune model: because $T_e$ and $T_i$ predictions in the core are the best validated among the transport channels, all outputs used in the cost function are masked except for $\rho<0.8$ for $T_e$ and $T_i$.}
    \label{fig:mask}
\end{figure}

Unfortunately, most models cannot predict more than a few quantities of interest with fidelity. Most egregiously, there is no good model for $Z_{eff}$, and in all simulations in this study it is held fixed at its initial value throughout the simulation as the best guess possible without giving information from the future. By contrast, machine learning models can attempt to learn arbitrary quantities, and in the worst case (assuming a perfect training process) can simply learn to e.g. hold $Z_{eff}$ constant if that is the best option for the training dataset. Therefore, when training the model with simulation outputs in the cost function as if they were experimental truth, a mask must be used so the model need not consider unpredicted quantities in the cost. In particular, the mask shown in Figure~\ref{fig:mask} is used to set as zero the portion of the cost function due to quantities which are not predicted well. In this study, because temperature predictions in the core are the best validated simulations, all profiles except $T_e$ and $T_i$, and all points $\rho>0.8$, have their cost function contributions fixed to zero. 

\begin{figure}
    \centering
    \includegraphics[width=0.5\textwidth]{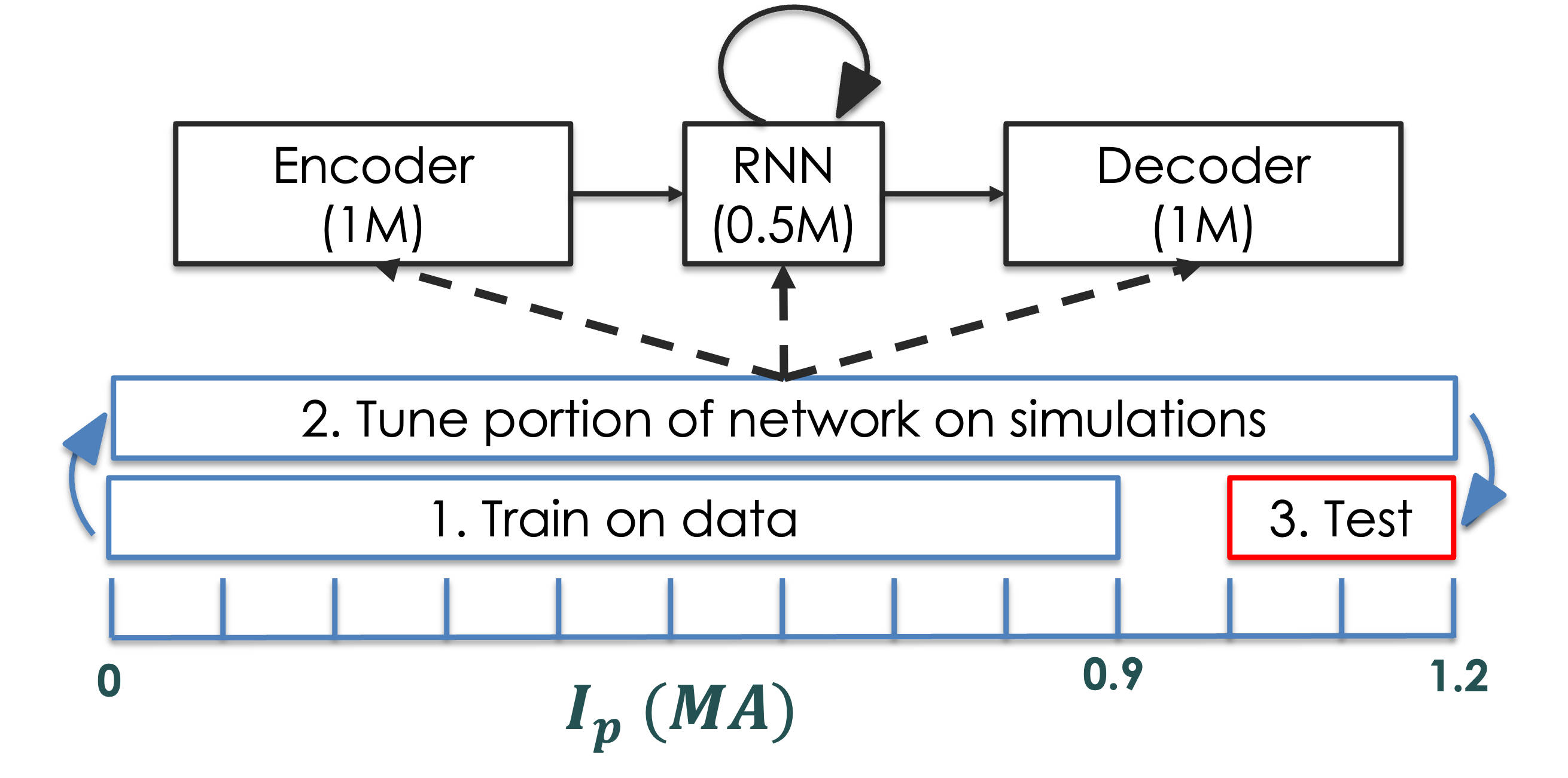}
    \caption{In transfer learning, the in-distribution dataset (in this case $I_p$ 0 to 0.9MA) is used to train a data-driven model. Next, a simulation database on the full distribution ($I_p$ 0 to 1.2MA) is created. The data-driven network is then tuned on this synthetic dataset, further training some or all of the architecture which includes 1 million weights in an encoder followed by 0.5 million weights in a recurrent network followed by 1 million weights in a decoder. The tuned network is tested out-of-distribution ($I_p$ from 1.0 to 1.2MA).}
    \label{fig:cartoon_transfer_learning}
\end{figure}

Note that training the model exclusively with simulation data would be equivalent to generating a surrogate model for the simulators. The goal here is instead to force the model to predict both simulation data and experimental data, in the hope that it generalizes better than experimental data only but is more accurate than simulation data only. Two mechanisms are used to help regularize the surrogate approach into a ``hybrid surrogate" approach. First, the model is tuned on near-distribution simulations \textit{alongside} the in-distribution experimental data rather than exclusively on simulations. Second, as an experiment, various portions of the model can be frozen (i.e. not trained) so that the model parameters cannot converge as far from the experimental values. This study compares the cases of no freezing, freezing just the encoder and decoder modules, and freezing just the recurrent module. A model is once again trained for $I_p$<0.9MA, and validated for 1.0<$I_p$<1.2MA, as shown in Figure~\ref{fig:cartoon_transfer_learning}.

\begin{figure}
    \centering
    \includegraphics[width=0.5\textwidth]{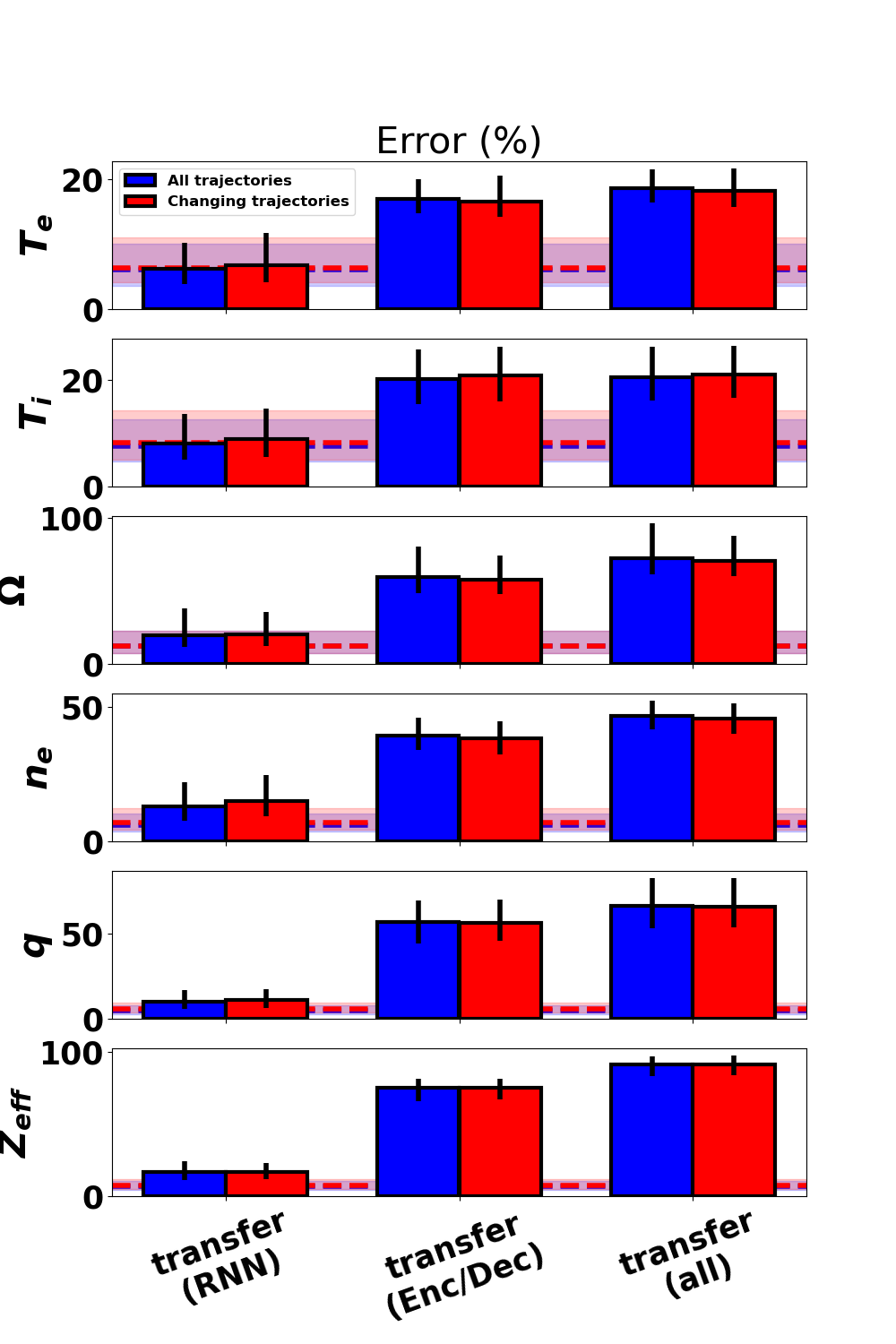}
    \caption{Transfer learning only degrades performance of the tuned models relative to data alone (horizontal dashed lines). Performance degrades less with fewer parameters allowed to be tuned, as demonstrated by the just-RNN-tuned model having the relative best performance (left) followed by the just-encoders-tuned model (model) followed by the model in which all 2.5 million weights are turned.}
    \label{fig:transfer_learning}
\end{figure}

It turns out this also yields no significant improvement over training on data alone, even for the case of extrapolating (to near-distribution) As shown in Figure~\ref{fig:transfer_learning}, transfer learning by bleeding in simulation data does worse whether it is the RNN, the encoders, or the full model that is tuned. Specifically, recall from Section~\ref{sec:machinelearning} the encoders collectively have about four times as many parameters as the RNN, so that the figure demonstrates that the further the model is allowed to veer from the initially learned model based on data only, the worse the predictions get.

\subsection{Concatenate simulation outputs to machine learning inputs}

In this section, a model is trained using outputs from simulations appended to the machine learning model's other inputs. Recall that at each timestep the machine learning model takes the actuators at the present and next timestep, along with the profiles (which may be from experiment or the machine learning model's own output from a previous timestep).

As described in Section~\ref{sec:physics}, there are a variety of calculations needed to determine sources and sinks for profiles prior to evolving the partial differential equations. These profiles of quantities, such as $S_\text{electron heat}$ (heat to electrons), $S_\text{ion heat}$ (heat to ions), $S_\text{current}$ (driven current), generally give a more detailed description of the way actuators interact with the plasma than the scalar controlled values (like $P_{NB}$ and $P_{ECH}$). We hypothesize giving the machine learning model estimates for these ``hidden parameters" from simulations may aid in training. 

As an alternative, the outputs of predictive simulations can also be concatenated as additional inputs to the machine learning model. As is general intuition in the field, the model may learn that simulations scale correctly but are known to over- or under-predict in known regimes. If the model learns the regimes, it can intelligently make the corrections. Because the heat transport is the most well-validated for tokamak predictions, $T_e$ and $T_i$ as predicted by TGLF-nn are considered as the additional inputs in this work. A model is once again trained for $I_p$<0.9MA, and validated for 1.0<$I_p$<1.2MA, as shown in Figure~\ref{fig:cartoon_transfer_learning}.

\begin{figure}
    \centering
    \includegraphics[width=0.5\textwidth]{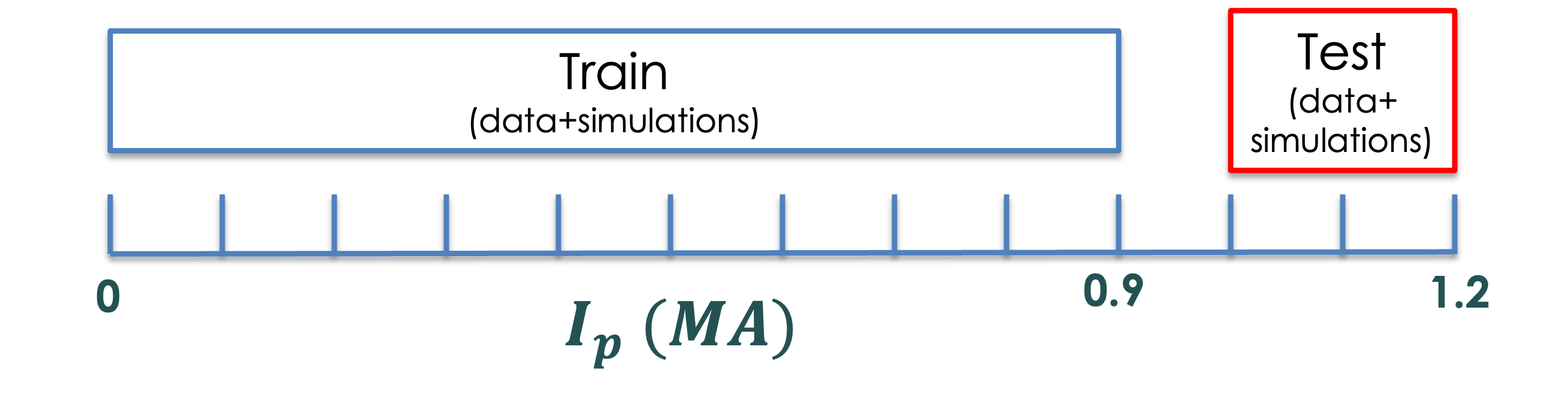}
    \caption{In concatenated context learning, the in-distribution dataset (in this case $I_p$ 0 to 0.9MA) is used to train a data-driven model, with additional inputs from simulations. The network is tested out-of-distribution ($I_p$ from 1.0 to 1.2MA), again with the added context from simulations.}
    \label{fig:cartoon_concatenation}
\end{figure}

\begin{figure}
    \centering
    \includegraphics[width=0.5\textwidth]{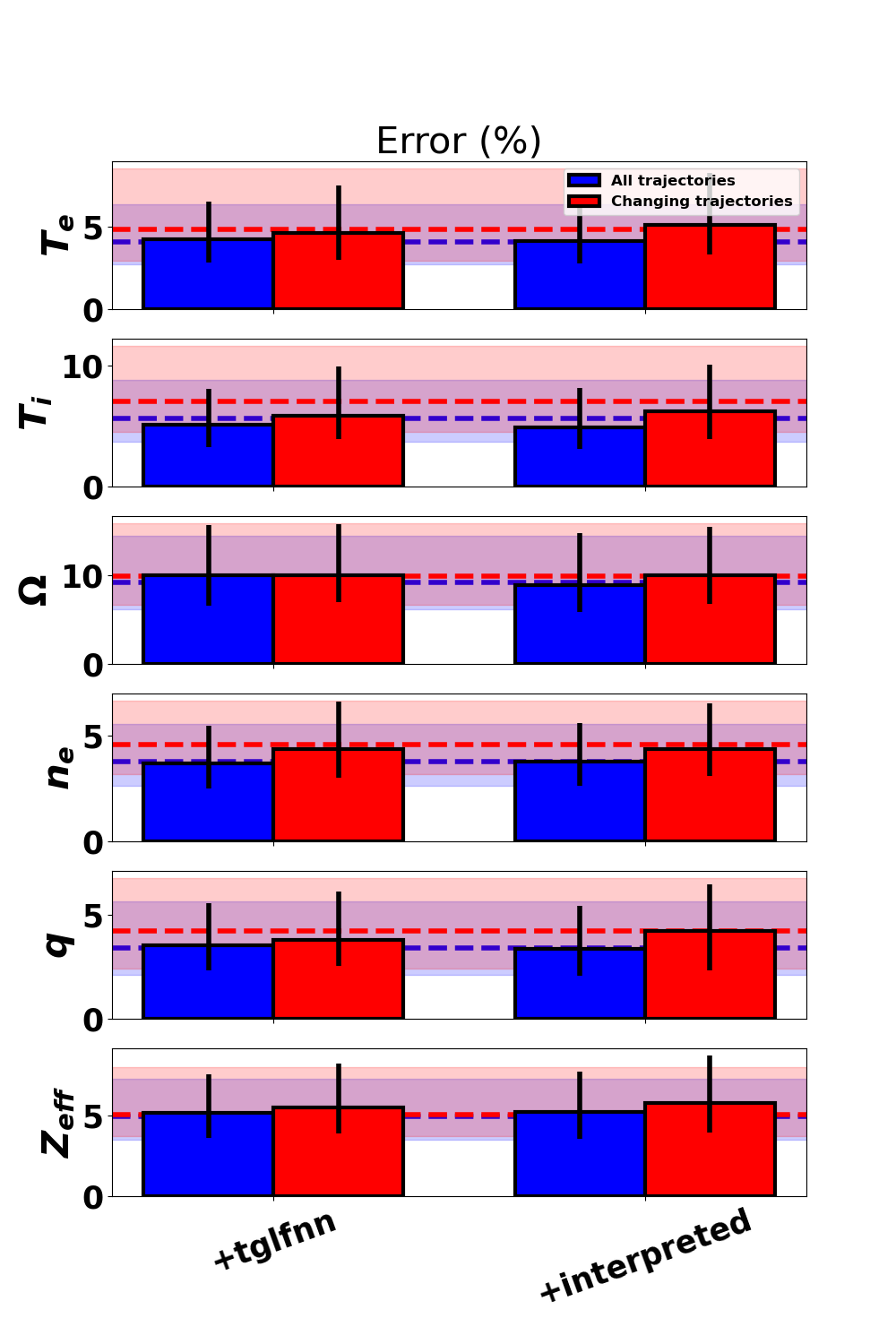}
    \caption{Concatenating context from simulations, including TGLF-nn diffusion coefficient profiles (left) and interpreted electron and ion heat and current drive profiles (right), yields little enhancement in predictions beyond data alone (dashed horizontal lines).}
    \label{fig:calculationsSigma}
\end{figure}

Once again, however, no significant improvement is achieved by adding these simulation outputs as inputs to the model (see Figure~\ref{fig:calculationsSigma}). Assuming the machine learning models were trained correctly, it appears that the physics simulators are not adding any information that cannot be learned by the models themselves, even for the purpose of extrapolating.

\subsection{Meta-learning}
\label{sec:stacked_generalization}
\begin{figure}
    \centering
    \includegraphics[width=0.5\textwidth]{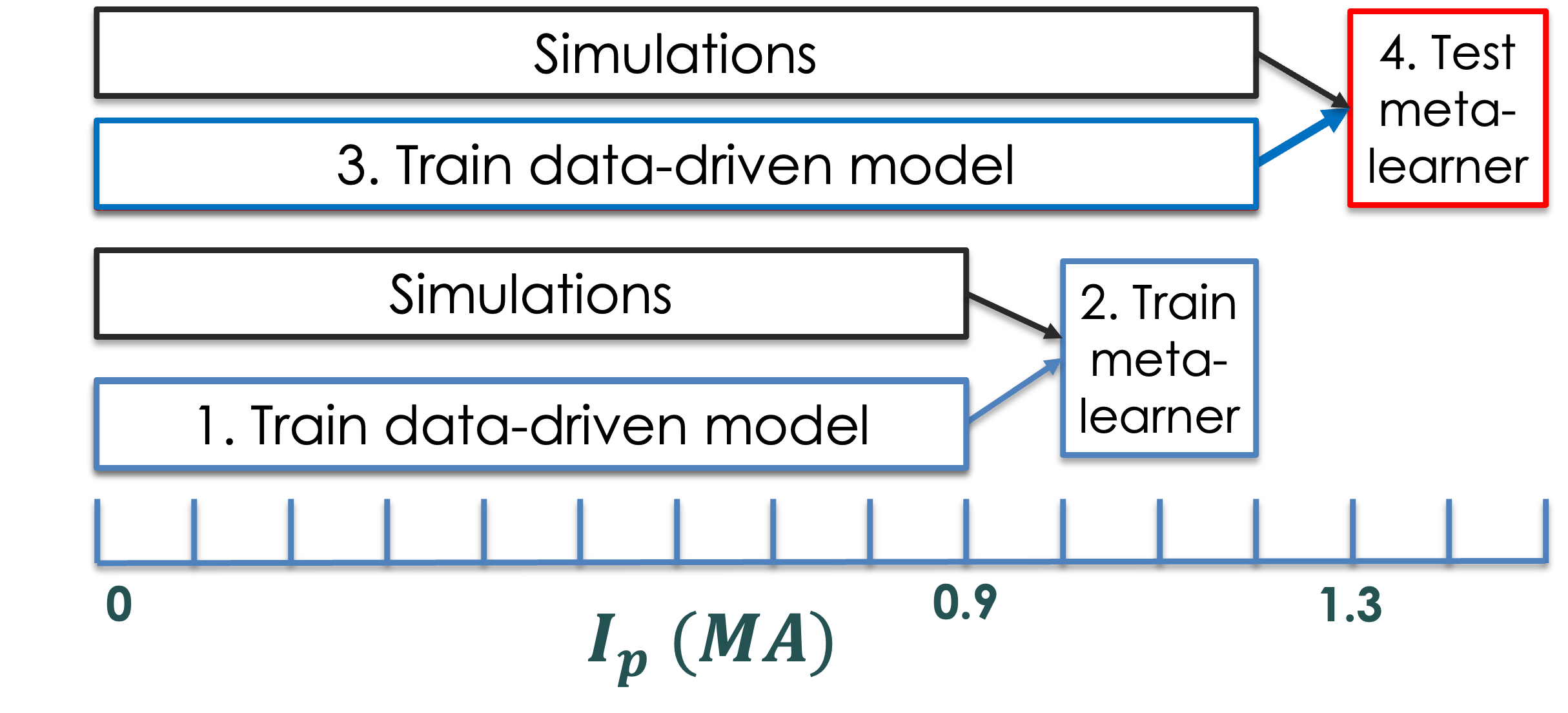}
    \caption{In meta-learning, a predictor is trained that takes as input the individual predictions from a variety of models (including physics-based simulators and data-driven models), effectively choosing how to weigh each of the predictions. The procedure for building such a model which extrapolates from in-distribution ($I_p$ 0 to 1.2MA in this case) to out-of-distribution ($I_p$>1.3MA) is to split the in-distribution data into two parts: a portion of data to train the data-driven models ($I_p$ 0 to 0.9MA) and a portion to train the meta-learner that extrapolates (to $I_p$ 1.0 to 1.2MA). The full dataset ($I_p$ 0 to 1.2MA) is then used to train a data-driven model, and finally the same meta-learner is used to extrapolate to the out-of-distribution data ($I_p$>1.3MA).}
    \label{fig:stacked_generalization_cartoon}
\end{figure}

A proposal in this section based on the concept of stacked generalization~\cite{wolpert_stacked_1992} should overcome some of the difficulties described in the previous section: in the case of simply adding simulation model outputs as machine learning inputs, there is no explicit reward (i.e. term in the cost function) for models extrapolating well to new regimes. We propose that should we want to build a predictor from $I_p$<1.2MA to predict $I_p$>1.3MA, we should first train a machine learning predictor for $I_p$<0.9MA. Then, a meta-learned model should be trained on 1.0MA<$I_p$<1.2MA, accepting as inputs the (extrapolated) output of the $I_p$<0.9MA model along with the output of simulators. The model should effectively learn how much to weight an extrapolated machine learning model vs simulators. With the meta-learning model optimized, one then trains a machine learning model on the full dataset ($I_p$<1.2MA) and the meta-learning model is used to predict $I_p$>1.3MA with the new machine learning model as input. The fundamental assumption is that the task of extrapolating from $I_p$<0.9MA to 1.0MA<$I_p$<1.2MA is similar to the task of extrapolating from $I_p$<1.2MA to $I_p$>1.3MA, in the sense of knowing how much to trust each model for the task of extrapolating. This training process is depicted in Figure~\ref{fig:stacked_generalization_cartoon}. 

A simple baseline is to average together the outputs of all models (i.e. sum the outputs with weight coefficients equal to $\frac{1}{\text{\# of models}}$). Even this simple ensembling approach may improve performance over baseline predictions, if the prediction errors are sufficiently uncorrelated and unbiased. To give a sense for this, consider that by the law of large numbers if all predictions are drawn from the same distribution and are uncorrelated and unbiased, the average will converge to the correct answer with enough models. This also demonstrates that when possible, the models should be as different as possible so that different models might work better in different regimes. 

\begin{figure}
    \centering
    \includegraphics[width=0.5\textwidth]{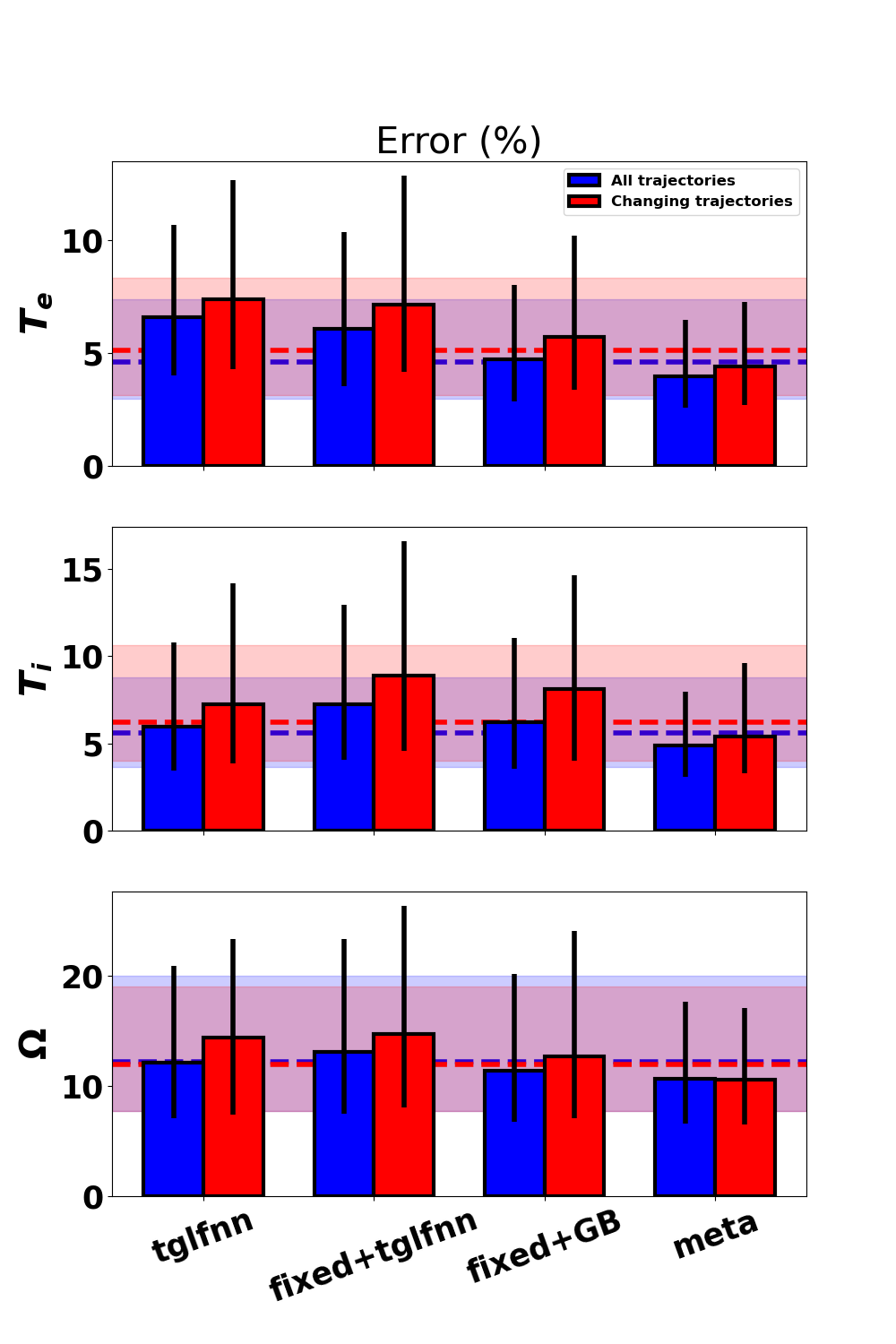}
    \caption{Meta-learning (right), in which a weighted average is taken of physics simulators and a data-driven model, yields a clear improvement in predictions beyond simulations (TGLF-nn, fixed-diffusion adjusted by TLGF-nn, and fixed-diffusion adjusted by a gyroBohm model) or data alone (dashed horizontal lines).}
    \label{fig:stacked_generalization}
\end{figure}

The architecture for the meta-learned model can take any form. As depicted in Figure~\ref{fig:mask}, a single sample from a given predictor has multiple profiles, multiple times, and multiple spatial locations. For the remainder of this paper, profile predictions are written as $X_\text{profile}^\text{predictor}(\rho,t)$ for different predictors ($n_\text{predictors}=4$: machine learning, ML; tglf-nn; fixed-tglf-nn; and fixed-gb), profiles ($n_\text{profiles}=3$: $T_e$, $T_i$, and $\Omega$), spatial locations $\rho$ ($n_\rho=33$, linearly between 0 and 1), and times $t$ ($n_\text{times}=15$, for 300ms sampled every 20ms). A learned model may consider a weighted sum of the $n_\text{predictors}$ different predictors, rather than a simple sum. Such a model must solve a logistic regression problem, i.e. learning scalar weights $\alpha_\text{predictor}$ corresponding to each predictor that sum to one,
\begin{align}
    \hat{X}_\text{profile}&=\sum_\text{predictors} \alpha_\text{predictor} X_\text{profile}^\text{predictor}(\rho,t) \\
    1 &= \sum_\text{predictors} \alpha_\text{predictor}.
\end{align}
PyTorch~\cite{paszke_pytorch_2019} is employed to solve the regression problem for the optimal $n_\text{predictors}-1$ learnable parameters $\alpha_i$, with the same loss as for the machine learning models (mean squared error on the normalized profiles). Similar to the case of transfer learning, only $T_e$, $T_i$, and $\Omega$ are ensembled to attempt to improve predictions since other transport channel predictions do not perform significantly better than assuming no change in profiles according to this study. As shown in Figure~\ref{fig:stacked_generalization}, the results from the optimized ensemble perform significantly better than the machine learning model trained on data alone ($I_p$<1.2MA), and surprisingly perform about as well as a machine learning model trained on all data. In other words, stacked generalization with simulations allows machine learning models trained in-distribution to perform on par with machine learning models trained near-distribution, even for a near-distribution test set.

\begin{figure}
    \centering
    \includegraphics[width=0.5\textwidth]{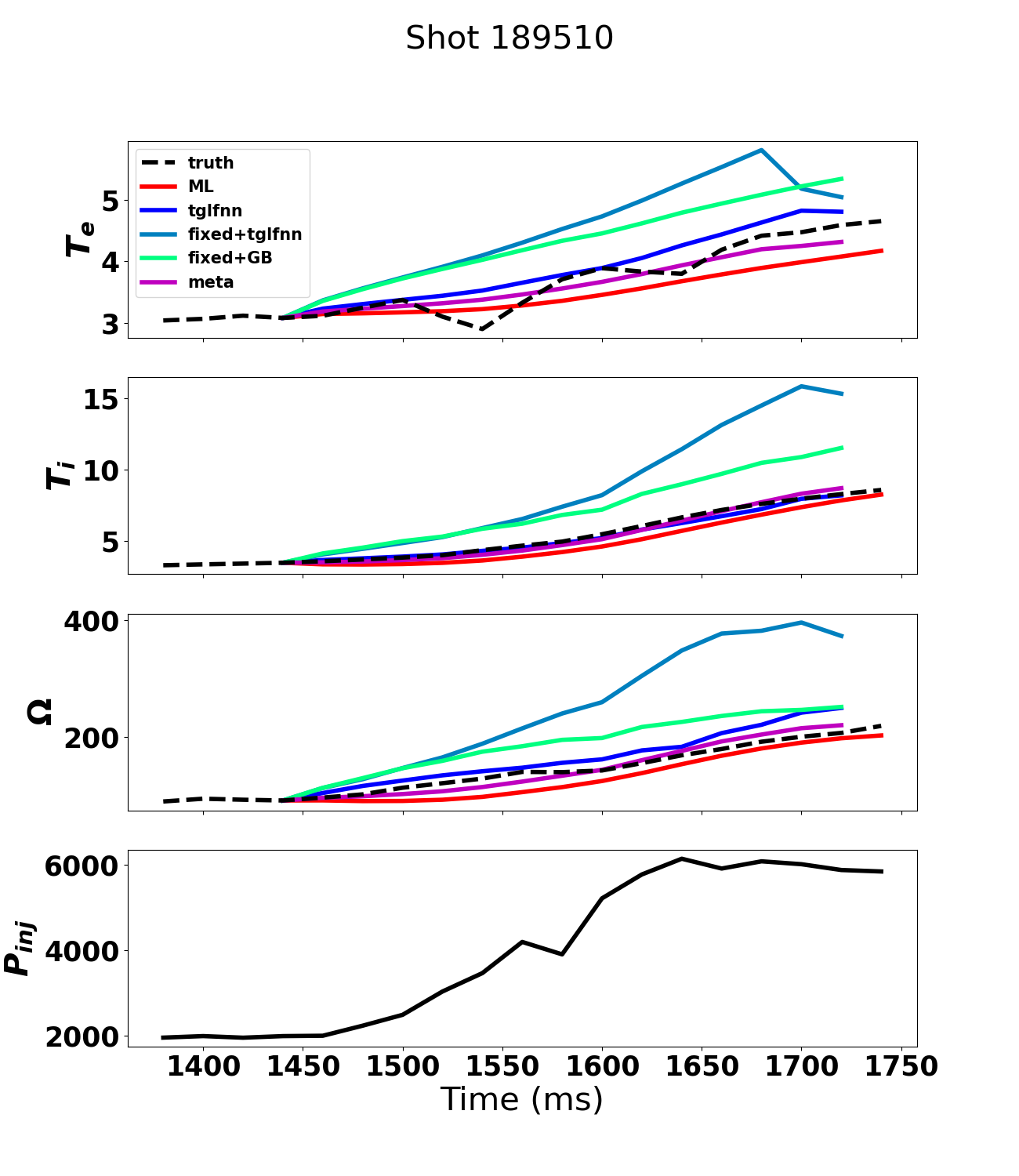}
    \caption{Timetrace of the core point for all model predictions vs true experimental measurement (dark black) illustrating how model errors can cancel one another out.}
    \label{fig:stacked_generalization_timetrace}
\end{figure}

\begin{figure}
    \centering
    \includegraphics[width=0.5\textwidth]{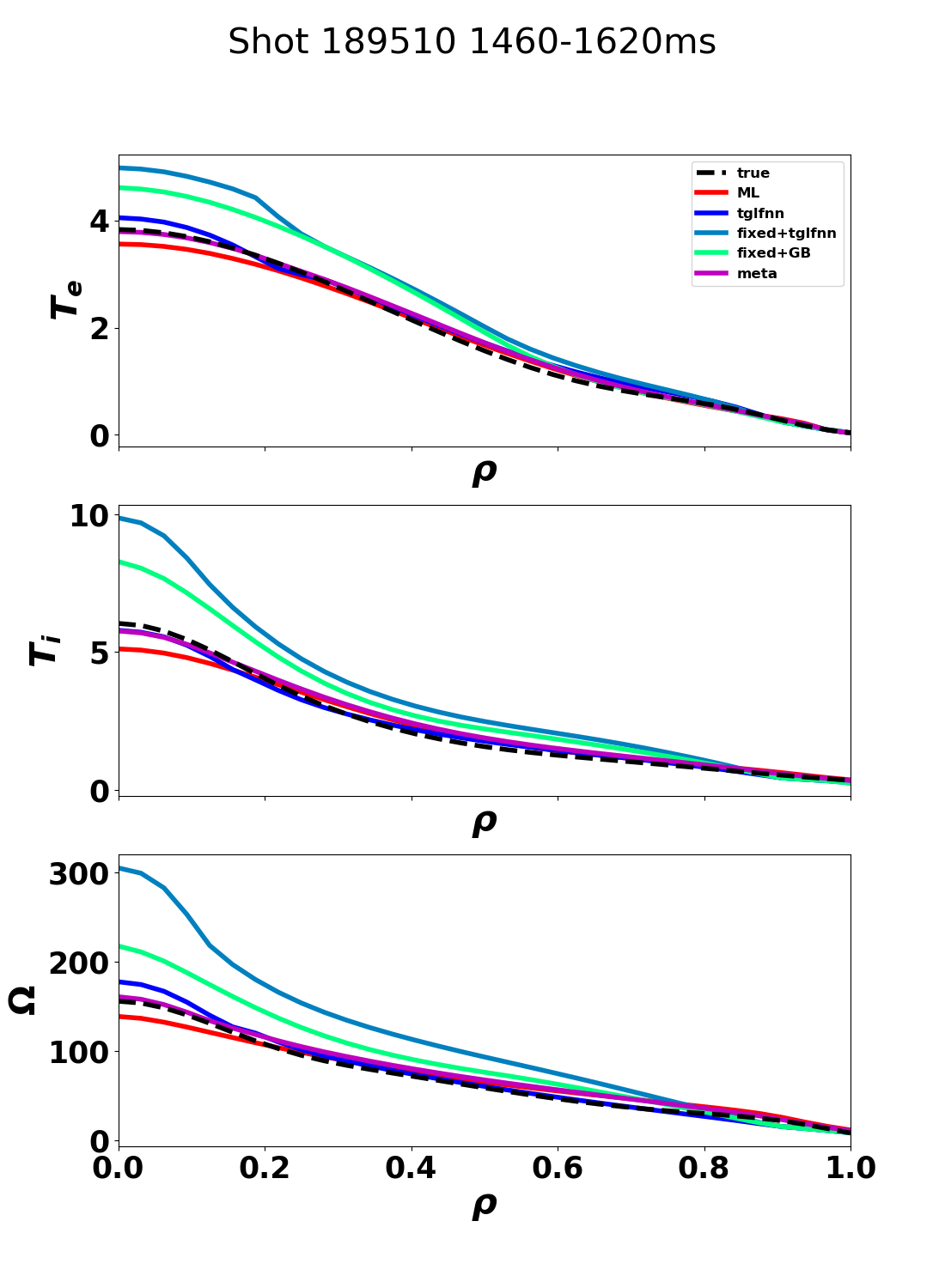}
    \caption{Timeslice of the models predictions between the initial time and a later timestep for the timetrace in Figure~\ref{fig:stacked_generalization_timetrace}. Note that the simulations in this case account for peaking of profiles in the core not captured by the machine learning model.}
    \label{fig:stacked_generalization_timeslice}
\end{figure}

The meta-learner chooses 64\% weight for the machine learning model, 19\% for the fixed-gyro-bohm model, 15\% for the TGLF-nn model, and 2\% for the fixed-TGLF-nn model. To give a sense of the mechanism in an example discharge (189510), Figures~\ref{fig:stacked_generalization_timetrace} and \ref{fig:stacked_generalization_timeslice} demonstrate the predictions over time of all the simulators, the $I_p$<1.2MA machine learning model, and the optimized ensemble predictor.

\subsection{Holistic confidence intervals}

All of these results can be viewed as confidence intervals for the improvement in quality of prediction over a baseline of a fully data-driven machine learning model. For simplicity, the errors are averaged over $T_e$ and $T_i$ such that a single number describes performance of a model. A 95\% confidence interval is employed.

\begin{figure}
    \centering
    \includegraphics[width=0.5\textwidth]{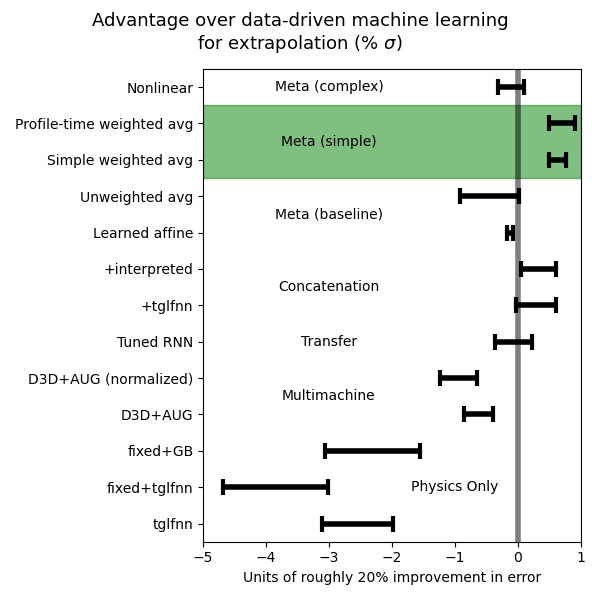}
    \caption{Confidence intervals for $\sigma$ percentage error improvement by each method beyond data-driven machine learning. Meta-learned models (green) yield a clear and significant improvement, with the half a percentage $\sigma$ improvement corresponding to a large (10\%) reduction in error for predictions, since typical $\sigma$ values are around 5\%.}
    \label{fig:holistic}
\end{figure}

Figure~\ref{fig:holistic} shows that among the models, meta-learning models have a clear statistical advantage over data-driven machine learning. Note from previous sections that $\sigma$ values typically fall around 5\% for $T_e$ and $T_i$, such that the half a percentage $\sigma$ reduction by the meta-learning models corresponds to a 10\% reduction in error beyond the data-driven machine learning model, as is clear also from Figure~\ref{fig:stacked_generalization}.

A few baselines are also shown in Figure~\ref{fig:holistic} to ensure the improved performance is statistically significant and owing to complementary information in simulations vs the machine learning models. One baseline is an unweighted average of the simulations with machine learning. Another is a learned affine model, i.e. linear regression learning scalar weight $\alpha$ and matrix offset $b_{\text{profile},\rho}$ such that
\begin{equation}
    \hat{X}_\text{profile}(\rho,t)= \alpha_\text{profile} X_\text{profile}^\text{ML}(\rho,t) + b_{\text{profile},\rho}.
\end{equation}
Note that this model therefore has $(1+n_\text{rho})n_\text{profiles}$ learnable parameters. As shown in Figure~\ref{fig:holistic}, these simpler models yield no performance improvement over data-driven machine learning.

For testing whether more complexity could improve the model, different learned weights $\alpha_{\text{predictor},\text{profile},\rho,t}$ for different profiles, for different points within each profile, and for each timestep are employed within the same meta-learned weighted sum framework as presented previously. Instead of simply taking a weighted sum of profiles for different models, this meta-learner takes a weighted sum of individual points within profiles. In other words,
\begin{align}
    \hat{X}_\text{profile}(\rho,t)&=\sum_{\text{predictors}} \alpha_{\text{predictor},\text{profile},\rho,t} X_\text{profile}^\text{predictor}(\rho,t) \\
    1 &= \sum_\text{predictors} \alpha_{\text{predictor},\text{profile},\rho,t}.
\end{align}
Note that this model effectively has $(n_\text{predictors}-1)n_\text{profiles}n_\text{rho}n_\text{times}$ learnable parameters. This ``Profile-time weighted avg" model, despite having many more learned parameters, does not perform much better than the simpler model.

An even more complex meta-learner can be built using a nonlinear model. Instead of the profile prediction being a weighted sum of each model's estimate, one can consider the profile prediction being a nonlinear transformation of each model's estimate. Analogous to the weighted sum, the transformation takes as input values from each of the considered models, and returns a single estimate for the true value. 
In this work, this nonlinear transformation is given by a multilayer perceptron. Specifically, the input vector is 1D (flattened) with values from each of $n_\text{profiles}$ profiles from each of $n_\text{predictors}$, followed by a matrix multiplication of $n_\text{predictors}n_\text{profiles}$ by $h$, for $h$ the dimension of the hidden nodes (5 in this work); followed by a rectified linear unit; followed by a matrix multiplication of $h$ by $n_\text{profiles}$ to give the needed output dimension. Using analogous notation to the linear models, with $h$ indexing the 5 hidden states; $M$, $M_\text{predictor,profile,h}$, $M'_\text{profile,h}$, $b_h$, and $b'_\text{profile}$ learnable scalars; and $v_h$ a dummy variable for ease of notation:
\begin{align}
    &v_h=\sum_\text{predictor,profile}
    M_\text{predictor,profile,h} X_\text{profile}^\text{predictor}(\rho,t)+b_h \\
    &\hat{X}_\text{profile}(\rho,t)=\sum_hM'_\text{profile,h} ReLU(v_h)+b_\text{profile}'.
\end{align}
For reference, the total number of learned parameters in the model is therefore order of $n_\text{profiles}n_\text{predictors}h=60$, compared to $n_\text{predictors}=4$ for the simple weighted average and $n_\text{predictors}n_{rho}=132$ for the model with different weights for each spatial location. In this particular "Nonlinear" model, performance drops well below the simpler meta-learners.

There is an infinite combination of different architectures one could consider, but the intuition to gather is that a simple model (weighted sum) is enough to yield performance improvement over simpler baselines such as unweighted sum; and that more complicated models do not yield much better performance if any.




\section{Discussion}
A meta-learning model was presented whose task is to extrapolate to new regimes given an ensemble of predictions. It can provide better extrapolated predictions than empirical or physics-based models alone. A model such as this that can extrapolate to new regimes while maintaining accuracy is broadly applicable to problems in engineering-oriented physics, such as for commissioning of reactor-scale fusion devices like ITER. 

This work also demonstrated methodologies for using data from one device ({AUG}) to extrapolate predictions of the state dynamics of another ({DIII-D}), which proved totally unhelpful for the specific example presented. One explanation could be that the physics on one device is so fundamentally different from that on another that new experiments like ITER will primarily rely on models trained on ITER itself, and not earlier devices. Another could be that profile fits are low fidelity and that uncertainty was not inherently included in the model. Future work might address each of these challenges by including probabilistic outputs (as in \cite{char_offline_2023}) as well as inputs to the model rather than point estimates, and using diagnostics directly in the cost function rather than relying on profile fits (i.e. calculating synthetic diagnostics from the predicted profiles during training). This would provide the additional benefit of augmenting the dataset with experiments that do not have clean profile fits and were therefore excluded in the training set of this study. One might also combine methodologies used in this paper, like adding physics simulation context for the case of multi-machine predictions. 

This work also demonstrated methodologies for transfer learning from simulations on never-before-seen discharges, which also demonstrated no performance improvements. However, this may be because there were a limited number of ``real" discharges the simulations were run on, so that the true dynamics of the simulator could not be learned. Future work may entail running simulations across a much larger quantity of ``fake" discharges, i.e. running the simulators with more arbitrary initial conditions and actuator trajectories, to fully leverage the benefits of transfer learning. 

\section*{Acknowledgements}
Thanks to David Eldon of the DIII-D team for extensive feedback on wording, clarification, and presentation of results.

This material is based upon work supported by the U.S. Department of Energy, Office of Science, Office of Fusion Energy Sciences, using the DIII-D National Fusion Facility, a DOE Office of Science user facility, under Award(s) DE-FC02-04ER54698; and PPPL contract DE-AC02-09CH11466.

\small{This report was prepared as an account of work sponsored by an agency of the United States Government. Neither the United States Government nor any agency thereof, nor any of their employees, makes any warranty, express or implied, or assumes any legal liability or responsibility for the accuracy, completeness, or usefulness of any information, apparatus, product, or process disclosed, or represents that its use would not infringe privately owned rights. Reference herein to any specific commercial product, process, or service by trade name, trademark, manufacturer, or otherwise does not necessarily constitute or imply its endorsement, recommendation, or favoring by the United States Government or any agency thereof. The views and opinions of authors expressed herein do not necessarily state or reflect those of the United States Government or any agency thereof.}

\bibliography{JoeDataSimPaper}

\end{document}